\documentclass[aps,reprint,twocolumn,footinbib,showpacs,superscriptaddress,longbibliography,pre]{revtex4-1}
\usepackage{graphicx}
\usepackage{dcolumn}
\usepackage{bm}
\usepackage{amssymb}
\usepackage{pifont}
\usepackage{amsmath}
\usepackage{times}
\usepackage[nooneline]{subfigure}

\usepackage{graphics}
\usepackage{graphicx}
\usepackage{xcolor}
\usepackage{float}
\usepackage{xr}
\usepackage{algorithm}
\usepackage{algpseudocode}
\newcommand{\PComment}[2]{\Comment{\parbox[t]{#1\linewidth}{\raggedright#2}}}

\begin{document}

\title{Synchrony-induced modes of oscillation of a neural field model}

\author{Jose M. Esnaola-Acebes}
\affiliation{Center for Brain and Cognition. Department of Information and Communication Technologies,
Universitat Pompeu Fabra, 08018 Barcelona, Spain}
\author{Alex Roxin}
\affiliation{Centre de Recerca Matem\`atica, Campus de Bellaterra, Edifici C, 08193 
Bellaterra (Barcelona), Spain.}
\author{Daniele Avitabile}
\affiliation{Centre for Mathematical Medicine and Biology, School of Mathematical Sciences, University of Nottingham, Nottingham, NG2 7RD, UK}
\author{Ernest Montbri\'o}
\affiliation{Center for Brain and Cognition. Department of Information and Communication Technologies,
Universitat Pompeu Fabra, 08018 Barcelona, Spain}
\date{\today}

\begin{abstract}
We investigate the modes of oscillation of heterogeneous ring-networks of quadratic
integrate-and-fire (QIF) neurons with non-local, space-dependent coupling. 
Perturbations of the equilibrium state 
with a particular wave number produce transient standing 
waves with a specific frequency, analogous to those in a tense string. 
In the neuronal network, the equilibrium  corresponds to a spatially homogeneous, 
asynchronous state. Perturbations of this state excite the network's 
oscillatory modes, which reflect the interplay of episodes of 
synchronous spiking with the excitatory-inhibitory spatial interactions. 
In the thermodynamic limit, an exact low-dimensional neural field model (QIF-NFM)
describing the macroscopic dynamics of the network is derived. 
This allows us to obtain formulas for the Turing eigenvalues of the
spatially-homogeneous state, and hence to obtain its stability boundary.
We find that the frequency of each Turing mode depends on the corresponding Fourier
coefficient of the synaptic pattern of connectivity. The decay rate instead, is
identical for all oscillation modes as a consequence of the
heterogeneity-induced desynchronization of the neurons. 
Finally, we numerically compute the spectrum of spatially-inhomogeneous
solutions branching from the Turing bifurcation, showing that similar oscillatory
modes operate in neural bump states, and are maintained away from onset.
\end{abstract}


\maketitle 

\section{Introduction}

Since the pioneering work of Wilson-Cowan~\cite{WC73}, Amari~\cite{Ama74,Ama77},
and Nunez~\cite{Nun74}, continuum descriptions of neuronal activity 
have become a powerful modeling  tool in neuroscience
~\cite{Erm98,Coo05,ET10,Bre12,CGP14,DJR+08}. 
Given that the number of neurons in a small region of cortex 
is very large, these descriptions consider neurons to be distributed along a 
continuous spatial variable, and the macroscopic state of the network to be 
described by a single, space-dependent, firing rate variable. 
The resulting neural field model (NFM) generally has the form of a continuous 
first order integro-differential equation, greatly facilitating 
the computational and mathematical analysis of the dynamics of large 
neuronal networks.

NFMs do not generally represent proper mathematical reductions 
of the mean activity of a network of spiking neurons. 
Nevertheless, NFMs
have proven to be remarkably accurate in qualitatively capturing the main types of
dynamical states seen in networks of large numbers of asynchronous
spiking neurons. For example it is well known that, in local 
networks of spiking neurons, differences between excitatory and inhibitory neurons 
can lead to oscillations~\cite{WC72,Erm94,BW03}. 
The generation of these oscillations does not depend on the spatial 
character of the network, and hence can be observed in non-spatially dependent 
firing rate models~\cite{Erm94}.
When the pattern of synaptic connectivity depends on the distance between neurons, 
NFMs show that these differences between excitation 
and inhibition can lead to the emergence of oscillations and waves
~\cite{Ama77,PE01a}.  Similar patterns can also be found in NFMs with   
spatially dependent delays ---modeling the effect of the finite velocity 
propagation of action potentials~\cite{WC73,JH97}---
as a great deal of theoretical work indicates, see e.g.
~\cite{CLO03,AH04,CL09,Zha07,Hut08,Tou12,Vel13,DGJ+15}. 

In some cases the spatio-temporal dynamics of NFMs has been directly compared 
to that observed in analogous networks of spiking neurons~\cite{RBH05,BBH07,RM11}.
In this work it was found that non-space-dependent delays predict the existence 
of many of the spatio-temporal patterns observed in asynchronous 
networks of spiking neurons with non-local, space-dependent interactions.
The success of NFMs in describing these patterns depends crucially on the 
spiking activity being highly asynchronous. In fact, it is well known that 
neural field descriptions fail to describe states characterized by a 
high degree of spike synchronization, see e.g~\cite{SOA13}.

Here we report a spatio-temporal dynamical feature of 
heterogeneous networks of spiking neurons with 
non-local interactions that, to the best of our knowledge, have been so far 
unexplored. We show that ring networks of spiking neurons 
display a number of discrete modes of oscillation, resembling those of a tense string. 
These modes are exclusively due to transient episodes of synchronous spiking 
and not due to the different 
time scales between excitation and inhibition, nor to the presence of any 
propagation or synaptic delay.

Traditional NFMs do not describe these synchrony-induced oscillations. 
Therefore, to investigate and characterize them, we 
apply a recent method to derive the firing rate equations 
of a globally coupled heterogeneous population of quadratic integrate and 
fire (QIF) neurons~\cite{MPR15}. This method, based on the so-called 
Ott-Antonsen theory~\cite{OA08,OA09,OHA11}, leads to an exact macroscopic 
description of the network in terms of two macroscopic variables: 
the mean firing rate and the mean membrane potential. 
The resulting mean-field model exactly describes any state of the system, 
including synchronous states. 
Here we extend the local firing rate model in~\cite{MPR15}, to include non-local, 
instantaneous interactions. The resulting neural field model 
for heterogeneous QIF neurons (QIF-NFM) clearly displays the synchrony-induced
oscillatory modes observed in simulations of spiking neurons.  
We then thoroughly investigate the QIF-NFM  
by means of both a linear and non-linear stability analysis of the 
spatially homogeneous state.
The analysis reveals the presence of an infinite number of 
oscillation modes, linked to the Fourier components of the spatial pattern of synaptic 
connections. The analysis also shows that all modes decay to the 
unpatterned state with the same rate, which depends on the 
degree of heterogeneity in the network. Finally, we investigate
the spectrum of the spatially inhomogeneous solutions 
of the QIF-NFM and find similar oscillatory modes also 
linked to transient episodes of spike synchronization.

\begin{figure}[t]
\includegraphics[width=\columnwidth,clip=true]{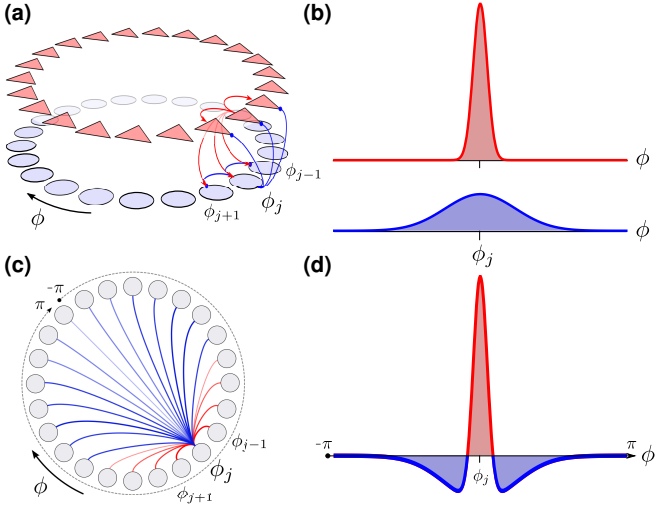}
\caption{(color online) Schematic representation of the ring network and 
coupling architecture under study. Panel (a) shows $N$
excitatory (red) and $N$ inhibitory (blue) neurons arranged on a ring. The location of 
neurons is parameterized by the angular variable 
$\phi_j = \frac{2 \pi j}{N}-\pi$, $j = 1, \dots, N$. Red and blue
lines indicate synaptic connections between neuron pairs $(\phi_j, \phi_k)$. 
An example of the excitatory and 
inhibitory space-dependent connectivity kernels Eqs~\eqref{J} 
are shown in panel (b) where the abscissa represents the distance,
$\left| \phi_k - \phi_j \right|$
between neurons $j$ and $k$. Panel (c) represents an effective model 
in which pairs of excitatory/inhibitory neurons located at a certain 
location $\phi_k$ are modeled as single 
neurons. The effective pattern of synaptic connectivity is obtained 
subtracting the inhibitory pattern from the excitatory one, as show in panel (d).}
\label{Figure1}
\end{figure}

\section{Synchrony-induced modes of oscillation in Networks of Quadratic Integrate 
and Fire (QIF) Neurons}

Figure~(\ref{Figure1}, a) shows a schematic representation of the spiking neuron network 
under investigation. The model consists of $N$
excitatory (Red) and $N$ inhibitory (Blue)
neurons evenly distributed in a ring, and characterized by the
spatial discrete variables  $\phi_j \in[-\pi,\pi)$ with
$\phi_j = \frac{2\pi j}{N} - \pi,\ j=1,\dots, N$, as shown in
Figure~(\ref{Figure1},a).
Any neuron in the network interacts with all the other neurons via the distance-dependent coupling function $J_{jk}^{e,i}=J^{e,i}(|\phi_j-\phi_k|)$, where
indices $e,i$ denote excitatory and inhibitory connections, respectively.
The synaptic projections of the $j$-th excitatory and inhibitory neurons (located at 
at $\phi_{j}$) to other two nearby neurons are also schematically represented in 
Figure (1,a). 

The ring architecture of the network allows one to express the 
excitatory and inhibitory connectivity patterns in Fourier series as  
\begin{equation}
J^{e,i}(\phi)= J_0^{e,i}+ 2\sum_{K=1}^\infty   J_K^{e,i} \cos (K \phi).\\
\label{J}
\end{equation}
Figure~(\ref{Figure1},b) shows a particular synaptic connectivity pattern in which excitatory  neurons form strong, 
short-range connections, whereas inhibitory
projections are weaker and wider. The state of the excitatory ($e$) 
and inhibitory ($i$) neurons is determined by the membrane 
potentials $\{v_j^{e,i}\}_{j=1,\ldots,N}$, which are modeled using 
the Quadratic Integrate and Fire (QIF) model~\cite{EK86,Izh07}
\begin{equation}
\tau \frac{dv_j^{e,i}}{dt}= (v_j^{e,i})^2 + I_j^{e,i},
\quad \text{(+ resetting rule)} .  
\label{qif1} 
\end{equation}
where $\tau$ is the cell's membrane time constant and, $v_r$ and $v_p$
correspond to the reset and peak potentials of the QIF neurons,
respectively ---in numerical simulations we consider $\tau=20$~ms.
The QIF neuron has two possible dynamical
regimes depending  on the input current $I^{e,i}_j$.
If $I^{e,i}_j<0$, the neuron is in the   
excitable regime, while for $I^{e,i}_j > 0$ the neuron is in the
oscillatory regime. In the excitable regime, an initial condition  
$v_j^{e,i}(0)<\sqrt{-I_j^{e,i}}$, asymptotically approaches   
the resting potential $-\sqrt{-I_j^{e,i}}$. On the other hand, 
initial conditions above the  
excitability threshold, $v_j^{e,i}(0)>\sqrt{-I_j^{e,i}}$, make the  
membrane potential to grow without bound. Specifically, if $v_j^{e,i}(0) \gg
\sqrt{I_j^{e,i}}$,  the membrane potential reaches infinity approximately 
after a time $\tau/v_j^{e,i}(0)$.
In practice, to avoid this divergence, we consider the following resetting rule:  
When the neuron's membrane potential  $v_j^{e,i}$ reaches a certain peak value 
$v_p \gg 1$, the neuron is reset to the new value 
$v_r=-v_p$ after a refractory period $2\tau/v_t$.
On the other hand, if $I_j^{e,i}>0$, the neuron is in the oscillatory regime and
needs to be reset periodically. If $v_p \gg 1$, the frequency of the oscillatory 
neurons is approximately $f_j=\sqrt{I_j}/(\tau \pi)$. 
Finally, the current $I_j^{e,i}$ is defined as 
\begin{equation}
I_j^{e,i}= \eta_j^{e,i} +\tau  S_j^{e}(t)+\tau S_j^{i}(t) + P_j^{e,i}(t).
\label{qif2}
\end{equation}
Here, $\eta_j^{e,i}$ is a constant external current, which varies 
from neuron to neuron 
(note that the voltages $v_j$ and currents $I_j$ are dimensionless). The terms $P^{e,i}(t)$ are time-varying common inputs, 
and $S_j^{e,i}(t)$ are the mean excitatory (positive) and inhibitory (negative) 
synaptic activities representing all the weighted inputs received by neuron $j$ 
due to spiking activity in the network:
\begin{equation}
S_j^{e,i}(t)=\pm  \sum_{k=1}^{N}  \frac{J_{j k}^{e,i} }{2\pi N}\sum_{l \backslash t_k^l<t} 
\frac{1}{\tau_s} \int_{t-\tau_s}^{t}dt' \delta^{e,i} (t'-t_k^{l}),
\label{s}
\end{equation}
where $\tau_s$ represents the synaptic processing time, 
and  $t_k^l$ is the time of the $l$-th spike of the excitatory/inhibitory $k$-th neuron. 

\begin{figure}[t]
\centerline{\includegraphics[width=\columnwidth,clip=true]{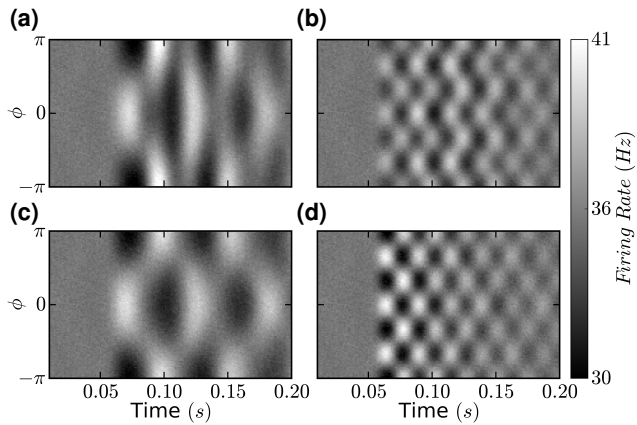}}
\caption[]{\label{Figure2} 
Transient episodes of spike synchrony in heterogeneous ring networks of 
$N =5\cdot 10^5$ QIF neurons, 
Eqs.~(\ref{qif1},\ref{qif2}), as a result of spatially-inhomogeneous 
perturbations applied  
at time $t=0.05$. In Panels (a,b) all excitatory neurons where perturbed. In
Panels (c,d) all neurons were perturbed. In panels (a,c) the 
perturbation had wavenumber $K=1$; in Panels (b,d) the perturbation had wavenumber $K=4$. 
Other parameters are: $\Delta = 1$, $\tau = 20$~ms, and $\bar\eta = 5$.
All Fourier components of the connectivity~Eq.~\eqref{J} were $J_K^{e,i}=0$, except:
$J^e_0=23$, $J^e_1=10$,  $J^e_2=7.5$,  $J^e_3=-2.5$,  $J^i_0=23$.}
\end{figure}

We performed numerical simulations of the QIF model 
Eqs.~(\ref{qif1},\ref{qif2}) for a network of 
heterogeneous neurons, see Figure \ref{Figure2}, and Appendix D 
for details of the numerical simulations. 
In all cases, the system is initially at a spatially homogeneous state (SHS). 
At time $t=50$~ms, a brief ($10$~ms) and small current pulse $P_j^{e}$  
is applied either to all excitatory neurons (panels a, b) or to both excitatory
and inhibitory neurons. 
Left and right panels show perturbations 
of the first spatial modes, respectively ---see Appendix D for the specific form of the perturbations. 
Note that, after the perturbation the system decays to the homogeneous 
state showing oscillations, which resemble standing waves. 
Note that the frequency of these oscillations is 
different for each mode, while the decay rate is similar in 
the two cases. We also performed simulations of networks of QIF neurons 
(i) with quenched Gaussian heterogeneity (ii) subject to independent Gaussian 
noise processes, and found similar results (not shown).
To the best of our knowledge, these oscillations 
have not yet been investigated in the literature.

\section{Neural Field model for quadratic integrate and fire neurons (QIF-NFM)}

In the following, we aim to investigate the nature and origin of the 
spatio-temporal patterns  shown in Figure~\ref{Figure2}. 
To analyze them, we derive the NFM corresponding to the 
thermodynamic ($N\to \infty$) and continuum limits of the network 
of QIF neurons Eqs.~(\ref{qif1},\ref{qif2}). 
In additon we also take the 
limit $v_p \to \infty$, so that the QIF model~\eqref{qif1} 
is equivalent to the so-called theta-neuron model~\cite{EK86,Izh07}. 
This leads to an exact neural field model for a network of QIF neurons (QIF-NFM)
\footnote{Invoking the Ott-Antonsen theory for populations of pulse-coupled 
theta neurons~\cite{LBS13,SLB14}, Laing recently derived a NFM~\cite{Lai14} that 
is equivalent to ours. However, in this work the resulting low-dimensional 
description is in terms of the complex Kuramoto order parameter. 
In contrast, the mean field description adopted here ---in terms of 
mean firing rates and membrane potentials--- greatly simplifies the 
analysis allowing us to analytically investigate the linear and 
non-linear stability of the spatially homogeneous states of the QIF-NFM.}.
The detailed derivation is performed in Appendix A, and closely 
follows that of~\cite{MPR15}. 
The reduction in dimensionality is achieved considering that the currents 
$\eta^{e,i}$ ---which, after performing 
the thermodynamic limit become 
continuous random variables--- are distributed according 
to a Lorentzian distribution of half-width $\Delta$ and centered at $\bar \eta$, 
\begin{equation}
g(\eta^{e,i})=\frac{\Delta}{\pi} \frac{1}{(\eta^{e,i}-\bar\eta)^2+\Delta^2}.
\label{g}
\end{equation}
The QIF-NFM is
\begin{subequations}
\label{freG}
\begin{eqnarray}
  \tau \frac{\partial R^{e,i}}{\partial t} &=& \frac{ \Delta }{\pi \tau}   + 2  R^{e,i} V^{e,i}  , 
\label{fre1G}\\
\nonumber
\tau \frac{\partial V^{e,i}}{\partial t} & =& (V^{e,i})^2+ \bar\eta-(\pi \tau
R^{e,i})^2+ \tau S(\phi) \\ & & +P^{e,i}(\phi,t).
\label{fre2G}
\end{eqnarray}
\end{subequations}
and exactly describes the time evolution of the mean 
firing rate $R^{e,i}(\phi)$, and the population's mean membrane potential 
$V^{e,i}(\phi)$ of the excitatory and inhibitory populations  
at any location $\phi$ of the ring ---to facilitate 
the notation we have avoided explicitly writing the dependence of these variables 
on $\phi$.
In the limit of instantaneous synapses, $\tau_s \to 0$ in Eqs.~\eqref{s}, 
the excitatory and inhibitory contributions of the mean field 
$S(\phi)=S^{e}(\phi)+S^{i}(\phi)$  
reduce to $S^{e,i}(\phi)=\pm \tfrac{1} {2\pi} \int_{-\pi}^{\pi} J^{e,i}(\phi-\phi')  
R^{e,i} (\phi') d \phi'$. 

\subsection{Effective QIF-NFM}
\label{sec:effective-qif-nfm}

The analysis of the QIF-NFM Eq.~\eqref{freG} is greatly simplified considering 
that excitatory and inhibitory neurons have identical 
single cell properties. This scenario is schematically represented in Figure~(\ref{Figure1};c,d). In this case, the solutions of Eqs.~\eqref{freG} satisfy  
$R^e(\phi,t)=R^i(\phi,t)\equiv R(\phi,t)$ and $V^e(\phi,t)=V^i(\phi,t)\equiv V(\phi,t)$. 
These solutions exist if $P^e(\phi,t)=P^i(\phi,t)=P(\phi,t)$, and coincide with the 
solutions of the  effective QIF-NFM 
\begin{subequations}
\label{fre}
\begin{eqnarray}
\tau \frac{\partial R}{\partial t} &=& \frac{ \Delta }{\pi \tau}   + 2  R  V  , 
\label{fre1}\\
\tau \frac{\partial V}{\partial t} & =& V^2+ \bar\eta-(\pi \tau R)^2+ \tau S(\phi) +P(\phi,t).
\label{fre2}
\end{eqnarray}
\end{subequations}
In this case, the mean field reduces to     
\begin{equation}
S(\phi)=\frac{1} {2\pi} \int\limits_{-\pi}^{\pi} \left[J_0
+2 \sum_{K=1}^\infty   J_K \cos (K (\phi'-\phi))\right] R(\phi') d \phi',
\label{mf}
\end{equation}
with the new Fourier coefficients $J_K$, which are related to those in 
Eq.~\eqref{J} as $J_K=J_K^e-J_K^i$, with $~K=0,1,\dots$, see Fig.\ref{Figure1}(d).
Note that, in Figs.~(\ref{Figure2};a, b), we perturbed the spatially homogeneous state (SHS) 
of the system Eqs.~(\ref{qif1},\ref{qif2}) using a current pulse to all 
the excitatory neurons. 
The resulting dynamics is only captured by the full system Eqs.~\eqref{freG}
and not by the effective neural field Eqs.~\eqref{fre}. 
However we next show that 
the existence of the spatial oscillatory modes observed in Fig.\ref{Figure2}
is exclusively linked to the dynamics in the 
reduced manifold defined by Eqs.~(\ref{fre}, \ref{mf}).

\subsection{Spatially homogeneous states (SHS) and their stability.
Synchrony-Induced Modes of oscillation}

In the following we investigate the stability of the stationary, 
spatially homogeneous states (SHS) of the QIF-NFM 
against spatial perturbations. The detailed linear stability 
analysis of both the complete model~\eqref{freG}, 
and the reduced one Eqs.~\eqref{fre} are provided in Appendix B. 

In absence of external 
inputs,  $P(\phi,t)=0$,  the steady states of Eqs.~\eqref{fre} ---and also of  
Eqs.~\eqref{freG}---, satisfy 
$V_*(\phi)=-\Delta/[2 \pi   \tau R_*(\phi)]$, and
\begin{equation}
R_*(\phi)=\Phi \left( \bar\eta +\tau S_*(\phi) \right)
\label{fpr}
\end{equation}
with $\Phi (x) = \sqrt{x +\sqrt{x^2+\Delta^2}}/(\sqrt{2}\pi \tau)$. 
In Eq.~\eqref{fpr}, the term $S_*(\phi)$ is 
the mean field Eq.~\eqref{mf} evaluated at $R_*(\phi)$.
For SHS, the mean field Eq.~\eqref{mf}
becomes spatially independent, $S_*(\phi)=S_*=J_0 R_*$, 
and Eq.~\eqref{fpr} becomes a quartic equation for the variable $R_*$.
To further simplify the analysis, hereafter we consider 
parameter ranges where Eq.~\eqref{fpr} has a single positive root. 
Accordingly, 
we consider a balanced kernel,
$J_0=0$ so that Eq.~\eqref{fpr} has $S_*=0$ and 
explicitly determines the value of the fixed point $R_*$.
%

\begin{figure}[t]
\includegraphics[width=\columnwidth]{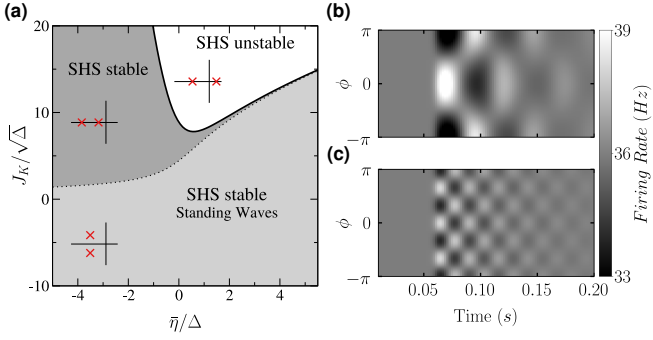}
\caption{(color online) (a) Phase diagram of Eqs.~\eqref{fre} (with $J_0=0$) 
showing the regions of stability of the Spatially Homogeneous State (SHS), 
determined by the eigenvalues Eq.~\eqref{lambda}.  
Spatial perturbations of wavenumber $K>0$ show oscillatory and non-oscillatory decay to the 
spatially homogeneous state in the light-shaded and dark-shaded 
regions of the diagram, respectively. 
The eigenvalues $\lambda_{K\pm}$ associated with the $K$-th mode
 are schematically represented in the complex plane
(Red crosses), 
for the three qualitatively different regions of the phase diagram.
Right panels show the 
response of the Eqs.~\eqref{fre} with $J_1=10$, $J_2=7.5$, $J_3=-2.5$ and 
$J_K=0$ ($K\neq1,2,3$), $\bar \eta=4.5$, $\Delta=1$ and $\tau=20$~ms, 
to a perturbation of the (b) $K=1$ and (c) $K=3$ spatial modes. 
Both perturbations produce standing waves 
with frequency and decay rate described by Eqs~\eqref{lambda}.  
In the white region, limited by the curve Eq.~\eqref{Jc}, these perturbations 
grow and lead to a Bump State (BS) with $K$ bumps (see Figure \ref{Figure4}). }
\label{Figure3}
\end{figure}

The steady states of the SHS of Eq.\eqref{fre} coincide with those of a 
single population of neurons~\cite{MPR15}. However, the stability of the 
SHS of the QIF-NFM to inhomogeneous perturbations 
depends on the spatial character of the connectivity kernel Eq.~\eqref{J}. 
The linear stability analysis of the SHS 
gives 
a countably infinite set
of eigenvalues associated to the stability of 
perturbations 
with wavenumber $K$
~\footnote{
The stability analysis of the original Eqs.~\eqref{freG} gives 
two additional complex eigenvalues for each oscillatory mode $K$. 
These eigenvalues are degenerated and are associated to the oscillatory modes of the 
uncoupled neuronal system,  that is they coincide with \eqref{lambda} 
with $J^{e,i}_K=0$. Additionally,
due to the translational invariance of the SHS solutions, 
each of the eigenvalues~Eq.~\eqref{lambda} 
is two-times degenerated, corresponding to 
even and odd perturbations. See Appendix B for 
the detailed linear stability analysis of the QIF-NFM Eqs.~\eqref{freG}.}.
\begin{equation}
\lambda_{K\pm}= -\frac{\Delta}{\pi \tau^2 R_*} 
\pm 2 \pi R_* \sqrt{\frac{J_K}{2 \pi^2 \tau R_* }-1}, ~(K=0,1,2\dots)
\label{lambda}
\end{equation}
This equation is the main result of this work, and explains the 
synchronization patterns shown in Fig.~\ref{Figure2}. 
Note that the eigenvalues Eq.~\eqref{lambda} may be real or complex, 
indicating non-oscillatory or oscillatory dynamics of the evolution of 
perturbations of wavenumber $K$, respectively. 
In particular, perturbations of any given spatial mode $K$ are oscillatory 
if the condition $J_K<2 \pi^2 \tau R_*$ is fulfilled. 
Notably, all complex 
eigenvalues have the same decay rate to the SHS, since   
Re$(\lambda_{K\pm})=-\Delta/(\pi\tau^{2} R_*)$ for all of them.
Specifically, the decay rate is proportional to the degree of quenched heterogeneity 
$\Delta$. This reflects the fact that the decay in the oscillations is in fact 
a desynchronization mechanism due to the distribution of inputs that the cells receive.

Substituting Eq.~\eqref{fpr} with $J_0=0$ into Eq.~\eqref{lambda}, 
it is straightforward to find the boundary 
\begin{equation}
J_K^o= \sqrt{2}\pi \sqrt{\bar \eta+\sqrt{\bar \eta^2+\Delta^2 }},
\label{Jc}
\end{equation}
separating the parameter space into regions where standing waves of wavenumber-$K$
are, or are not observed. This boundary is depicted with a dotted line in the 
phase diagram Fig.~\eqref{Figure3}, 
together with a schematic representation of the location of the 
eigenvalues $\lambda_{K\pm}$
in the complex plane (red crosses, see also Fig.(\ref{Figure5},a)).

A given oscillatory mode $K$ has an associated frequency 
$\nu_K = 1/(2 \pi) |\text{Im}(\lambda_{K\pm})|$, which  
differs from one another depending on the corresponding Fourier 
coefficients $J_K$ of the 
patterns of synaptic connectivity Eq.~\eqref{J}. Therefore, spatial perturbations
of wavenumber $K$, produce standing waves of neural activity of frequency 
$\nu_K$. 
Locally excitatory coupling $J_K>0$ slows down these oscillations 
and eventually suppresses them, whereas locally inhibitory coefficients 
$J_K<0$ are able to generate 
arbitrarily fast oscillations  (in particular, note that all modes with $J_K=0$ 
are oscillatory with frequency $\nu= R_*$, which coincides with the 
mean firing rate of the uncoupled neurons).

Indeed, in Fig.~(\ref{Figure2},d), a perturbation of wavenumber $K=3$ produced standing waves, since $J_3$ was negative. The frequency of these oscillations was
fast compared to that of Fig.~(\ref{Figure2},c), where the 
exited mode was the first one $K=1$, and given that the $J_1$ was positive. 
However, note that in both cases the decay 
to the SHS is similar, as predicted by the eigenvalues Eq.~\eqref{lambda}. 
This indicates that the desynchronization process occurs faster when the 
diversity $\Delta$ of neurons is increased, 
and this process doesn't depend on the oscillation mode being excited. 
Finally, in panels (b,c) of Fig.~\ref{Figure3} we show 
numerical simulations of the QIF-NFM Eq.~\eqref{fre} using the same parameters as 
those of Fig.~\ref{Figure2} (c,d), and the agreement is good.

\subsection{Turing bifurcation and nonlinear stability of the SHS}

As $J_K$ is increased, the frequency $\nu_K$ of a given oscillatory 
mode decreases and eventually it ceases to oscillate. 
Further increases in $J_K$ may destabilize the homogeneous state, via a
pattern-forming (Turing) bifurcation. This instability leads to states 
with spatially modulated firing rate, sometimes referred to as 
Bump States (BS). Substituting the fixed point \eqref{fpr} in 
Eq.~\eqref{lambda}, and imposing the condition of marginal stability 
$\lambda_{K+}=0$, we find the stability boundaries corresponding to 
a $K$-spatial mode
\begin{equation}
J_{K}^T  =2\pi\sqrt{\frac{2 \bar \eta^2+2 \Delta^2}
{\bar \eta+ \sqrt{\bar \eta^2+\Delta^2 } }}.
\label{JT}
\end{equation}
The Turing bifurcation boundary, Eq.~\eqref{JT}, corresponds to the  
solid line in Figs.~(\ref{Figure3}a,\ref{Figure4}a). 
Additionally, in Appendix C, we conducted a weakly nonlinear 
analysis and derived the small amplitude equation Eq.~\eqref{AmplEq} 
corresponding to the 
bump solution bifurcating from the SHS.
The amplitude equations determine if the Turing bifurcation is supercritical,
or if it is subcritical and bistability between SHS and Bump states 
is expected to occur. The results of this analysis are summarized in 
Fig.(\ref{Figure4},b). 

In addition, we performed numerical simulations of the 
QIF-NFM~\eqref{fre}, and indeed found 
coexistence of SHS and Bump states in the blue-shaded
regions limited by solid and dashed curves in Fig.(\ref{Figure4},a). 
These lines meet at two codimension-2 points (where the Turing 
bifurcation line changes color) that agree with 
the results of the weakly nonlinear analysis. 
Moreover, we computed numerically a bifurcation diagram of the NFM,
  using the spectral method developed in Reference~\cite{RA14} and
  available with Reference~\cite{A16}. The results, presented in
  Figure (\ref{Figure4},c) confirm that the unstable BS  bifurcates subcritically for
  the SHS. The unstable BS then meets a stable BS ---solid Blue line--- at a fold
  bifurcation.   

%

\begin{figure}[t]
\centerline{\includegraphics[width=\columnwidth]{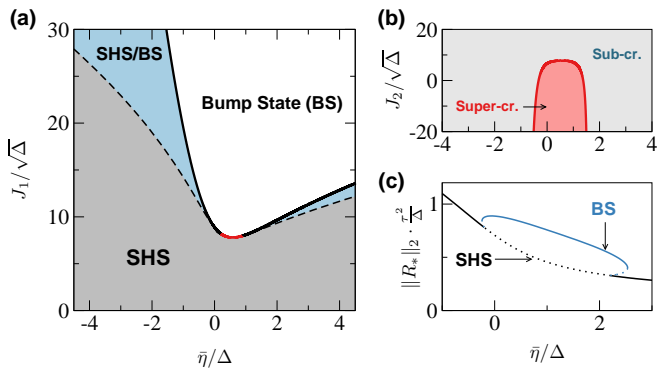}}
\caption{(color online) (a) Phase diagram of the 
QIF-NFM Eqs.~\eqref{fre} with $J_2=7.5$, $J_3=-2.5$, $J_K=0$ 
for $K>3$, and $\Delta=1$. Solid line: Supercritical (Red) and 
Subcritical (Black) Turing bifurcation boundary Eq.~\eqref{JT}.
Dashed lines: Saddle node bifurcation of bumps (numerical).  
(b) Diagram ---obtained using a weakly nonlinear analysis---
showing the regions where the Turing bifurcation 
is supercritical or subcritical, for $J_1=10$, 
$J_3=-2.5$, and $J_K=0$.  
(c) Bifurcation diagram (rescaled) 
$\|R_*\|_2=(2\pi)^{-1}\int_{-\pi}^{\pi}|R_*(\phi)|^2 d\phi$
vs. $\bar \eta$, for $J_1=10$.
Solid/Dotted Black lines: Stable/Unstable SHS.
Solid/Dotted Blue lines: Stable/Unstable Bump States (BS). }
\label{Figure4}
\end{figure}

\subsection{Synchrony-induced transient oscillations in Bump states}

\begin{figure}[t]
\centerline{\includegraphics[width=85mm,clip=true]{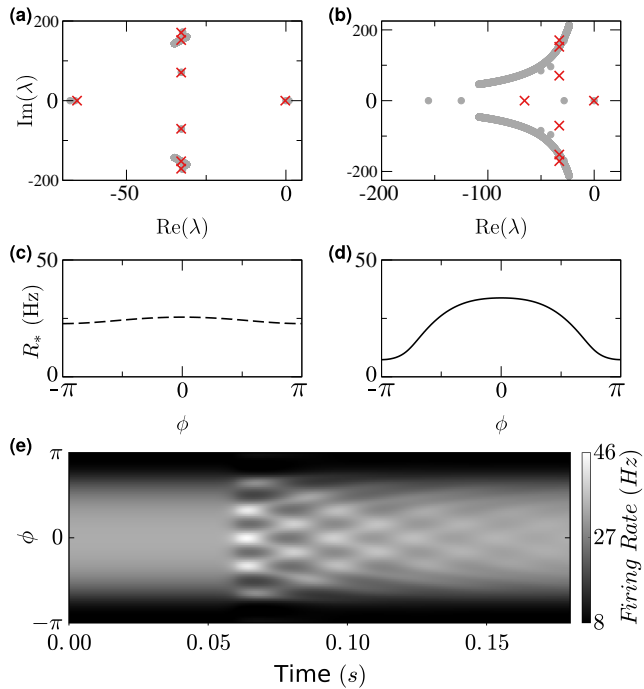}}
\caption{(color online)  Spectrum (a,b) and firing rate 
profiles (d,c) of an unstable (a,c) and
stable (b,d) Bump States of the QIF-NFM Eqs.~\eqref{fre}.
In panel (c) the eigenvalues Eq.~\eqref{lambda} are 
superimposed with red crosses. 
Panel (e) shows a numerical simulation of the BS of panel (d). At 
$t=0.05$~s, a perturbation of wavenumber $K=6$ is applied. 
Parameters are
$J_0=0$, $J_1=10$, $J_2=7.5$, $J_3=-2.5$, $J_K=0$ for $K>3$, $\Delta=1$,
$\tau=20$~ms. Panels (a,c): $\bar \eta=2.2120$; Panels (b,d,e): $\bar \eta=2.1828$.}
\label{Figure5}
\end{figure}

To investigate whether the synchrony-induced oscillatory modes are also present in the 
stationary BS, we computed their spectrum.
The gray points in Fig.~(\ref{Figure5},a) show the spectrum of the unstable Bump
near the subcritical Turing Bifurcation of wavelength $K=1$.
Additionally, the red crosses in Fig.~(\ref{Figure5},a) 
are the eigenvalues of the SHS state Eq.~\eqref{lambda}. 
The profile of the unstable bump is only very weakly modulated, 
see Fig.~(\ref{Figure5},c), and hence the spectrum of the BS is very close to 
that of the SHS, given by the eigenvalues $\lambda_K$. 
All these eigenvalues are complex, except two real eigenvalues which correspond 
to the $K=1$ mode. One of these eigenvalues is negative and 
the other is very close to zero and positive, indicating that the SHS is unstable.

Additionally, it is important to note that in Fig.~\ref{Figure5} we have 
taken $J_K=0$ for all $K$ except for $K=1,2,3$, and hence 
 there is an infinite number of eigenvalues ($\lambda_0$ and $\lambda_{4,5, \dots}$) 
that are all complex and identical. 
In Fig.~(\ref{Figure5},a) the eigenvalues of the unstable BS seem to form  
a continuous band precisely around these infinitely degenerated eigenvalues
and their complex conjugates. These continuous bands grow in
size as one moves away from the Turing bifurcation, as it can be seen
in the spectrum of the stable bump depicted in Fig.~(\ref{Figure5},b)
---here red crosses also correspond to the eigenvalues of the SHS 
state Eq.~\eqref{lambda}. 
These results show that all the complex eigenvalues linked to the
oscillatory modes of the SHS remain complex, suggesting that, in general, similar
synchronization-induced oscillations may be present in stationary, spatially
inhomogeneous neural patterns. 

Finally, to illustrate this, in Fig.~(\ref{Figure5},e) we performed a numerical 
simulation of the QIF-NFM Eqs.~\eqref{fre}, and perturbed the BS shown in 
Fig.(\ref{Figure5},d) with a spatially inhomogeneous 
perturbation corresponding to the mode ($K=6$). 
The perturbation decays to the BS showing a pattern that resembles that of 
Figs.~\eqref{Figure2}. However here, the regions of the ring with the maximum  
values of $R_*$ ---around $\phi=0$, in panels (d,e)--- oscillate 
at high frequencies
and these oscillations slow down 
as $\phi \to\pm \pi$.
The spectrum of the stable BS Fig.~(\ref{Figure5},b) also indicates that the 
decay of the fast oscillations (located at the central part of the bump, $\phi=0$) 
is slow compared to that of the slow oscillations.

\section{Conclusions}

We have reported the existence of a class of oscillatory modes in 
spatially distributed networks of heterogeneous 
spiking neurons. These modes of oscillation 
reflect the transient episodes of spike synchronization among the neurons 
and are not captured by traditional NFMs. To investigate them, we derived a novel 
NFM for QIF neurons, Eqs.~\eqref{freG} and \eqref{fre}, which 
allows us to find the eigenvalues determining the linear stability of the spatially 
homogeneous state. This analysis reveals  two important features: 
(i) The frequency of each oscillation mode only depends 
on the corresponding Fourier coefficient of the synaptic pattern of 
connectivity; (ii) The decay rate is exactly the same for all
modes, and is due to a desynchronization mechanism which  
depends on the degree of quenched heterogeneity. 
We also numerically investigated networks of identical QIF neurons subject to noise, 
and found similar results (not shown). In this case the desynchronization 
reflects an underlying phase diffusion proportional to the noise strength. 
Finally we investigated the existence and stability of bump states, 
which bifurcate from the spatially homogeneous 
states via Turing bifurcations. The spectrum of such bump states 
has a continuous part off the real axis, 
indicating that similar synchronization-induced oscillatory modes 
also operate in neural bump states.

Interesting directions of further study are the analysis of the 
QIF-NFM~\eqref{freG} considering different membrane time constants $\tau$,
(or different main currents $\bar \eta$) for excitatory and inhibitory neurons.
As proved recently~\cite{AD17}, NFMs with time-scale separation display a rich
variety of robust spatio-temporal patterns, which may also be supported by our
model.
Also, recent work has been done to extend the local firing rate 
equations derived in~\cite{MPR15} to include synaptic kinetics~
\cite{RP16,CB16,BBC17,DRM17} or fixed delays \cite{PM16}. These studies all show 
that time delays due to synaptic processing generally lead to the emergence of 
self-sustained oscillations due to collective 
synchronization. Extending the QIF-NFM~\eqref{freG} 
to account for the synaptic time delays caused by synaptic processing 
may lead to spatio-temporal phenomena not previously observed 
in traditional NFMs.

\begin{acknowledgments}
J.M.E.-A. and E.M acknowledge support 
by the European Union's Horizon 2020 research and innovation
programme under the Marie Sk{\l}odowska-Curie grant agreement No.~642563.
J.M.E.-A. and E.M. acknowledge the projects grants 
from the Spanish ministry of
Economics and Competitiveness, Grants No.~PSI2016-75688-P and 
No.~PCIN-2015-127.
A.R. acknowledges a project grant from the Spanish ministry of
Economics and Competitiveness, Grant No. BFU2012-33413.
A.R. has been partially funded by the CERCA progam of the
Generalitat de Catalunya. D.A. was partially supported by the EPSRC grant
EP/P510993/1 (United Kingdom).
\end{acknowledgments}


%

\newpage


 \makeatletter 
\setcounter{equation}{0} 
\renewcommand{\theequation}{A\arabic{equation}}
 \makeatother

\section*{Appendix A: Derivation of the QIF neural field model (QIF-NFM)} 
 
Our derivation closely follows that of~\cite{MPR15}, but it needs to
be extended to include the spatial dimension. 
Similar extensions from a single population of phase oscillators 
to a one dimensional, spatially distributed network with non-local 
coupling have been done 
in~\cite{Lai14,Lai15,Lai16i,Lai16ii,Lai09i,Ome13,OWL14,Kaw14}.  

Considering the thermodynamic limit $N\to\infty$, we can drop the 
indexes in Eqs.~(\ref{qif1}, \ref{qif2})  
and define the density function $\rho^{e,i}(v^{e,i} \vert \eta^{e,i},
t, \phi)$ such that $\rho^{e,i}(v^{e,i} \vert \eta^{e,i},
t, \phi)dv^{e,i}  d\eta^{e,i}  d\phi $  describes the fraction of neurons located  
between $\phi$  and $\phi + d\phi$, with membrane potentials between $v^{e,i} $ and 
$v^{e,i} +dv^{e,i} $, and parameters between $\eta^{e,i}$ and 
$\eta^{e,i} + d\eta^{e,i}$ at time $t$.  
Accordingly, parameter $\eta^{e,i}$ becomes now a continuous  
random variable with probability density function $g(\eta^{e,i})$. For
the sake of simplicity we assume identical  distributions
for both excitatory and inhibitory populations $g \left( \eta^{e,i}
\right) = g \left( \eta \right)$.  
The total voltage density at location 
$\phi$ and time $t$ is given by 
$\int_{-\infty}^{\infty} \rho^{e,i}(v^{e,i} \vert \eta, t, \phi)~g(\eta)~ d\eta$. 
 
Conservation of the number of neurons at each $\phi$ value  
is described by the continuity equation
$$\partial_t \rho^{e,i} =  
- \partial_{v} \left[  \left( \left( v^{e,i} \right)^2 +\eta + \tau S(\phi,t)+ P^{e,i}(\phi,t) \right) \rho^{e,i} \right],$$ 
where we have explicitly included the velocity given by  
equations \eqref{qif1} and \eqref{qif2} and $S \left( \phi,t \right) =
S^e \left( \phi,t \right) + S^i \left( \phi,t \right)$ represents the
total synaptic activity. 
Next we invoke the Ott-Antonsen theory~\cite{OA08},  
by means of the Lorentzian Ansatz (LA) \cite{MPR15}  
\begin{equation} 
  \rho^{e,i}(v^{e,i} \vert \eta, t, \phi)= \frac{1}{\pi}\frac{x^{e,i}(\phi,\eta,t)} 
  {\left[v^{e,i}-y^{e,i}(\phi,\eta,t)\right]^2+x^{e,i}(\phi,\eta,t)^2}, 
  \label{eq:la} 
\end{equation} 
which solves the continuity equation. 
The width $x^{e,i}(\phi,\eta,t)$ of the LA is related to the firing rate $R^{e,i}$
of the  
neural populations. Indeed, for each $\eta$ value at time $t$,  
$R^{e,i}(\phi,\eta,t)$ can be evaluated noting that neurons  
fire at a rate given by the probability flux at infinity:  
$R^{e,i}(\phi,\eta,t)= \rho^{e,i}(v^{e,i} \to \infty|\eta,t, \phi) \dot v^{e,i} (v^{e,i} \to
\infty|\eta,t, \phi).$ 
The limit $v^{e,i}\to\infty$ on the right hand side of this equation can be
evaluated within the LA, and gives: $x^{e,i}(\phi,\eta,t)=  \pi \tau
R^{e,i}(\phi,\eta,t)$. The total firing rate at a particular location
$\phi$ of the ring is then   
\begin{equation} 
  R^{e,i}(\phi,t)=\frac{1}{\tau
    \pi}\int_{-\infty}^{\infty}x^{e,i}(\phi,\eta,t)g(\eta)d\eta. 
  \label{eq:r} 
\end{equation} 
Additionally, the quantity $y^{e,i}(\eta,t)$ is, for each value of $\eta$, 
the mean of the membrane potential $ 
y^{e,i}(\phi,\eta,t)= \mathrm{P.V.}  
\int_{-\infty}^{\infty} \rho^{e,i}(v^{e,i} |\eta,t, \phi) v^{e,i} \, dv^{e,i}$. 
Therefore, this variable  is related to the  
mean membrane potential of the neuronal population at $\phi$ by  
\begin{equation} 
  V^{e,i}(\phi,t)= \int_{-\infty}^{\infty}  y^{e,i}(\phi,\eta,t)g(\eta)  d\eta. 
  \label{eq:v} 
\end{equation} 
Substituting the LA \eqref{eq:la} into the continuity equation,  
we find that, for each value of $\eta$, 
the variables $x^{e,i}(\phi)$ and $y^{e,i}(\phi)$ must 
obey two coupled equations which can be written in complex form as  
\begin{equation}
  \label{eq:w}
  \begin{split}
    \tau \partial_t w^{e,i}(\phi,\eta,t)= i \big[  \eta  
    &+ \tau  S(\phi,t)- \left( w^{e,i} \right)^2(\phi,\eta,t) \\
    & + P^{e,i}(\phi,t) \big],
  \end{split}
\end{equation}\\[0.3em]
where $w^{e,i}(\phi,\eta,t)\equiv x^{e,i}(\phi,\eta,t) +i y^{e,i}(\phi,\eta,t).$ 
If $\eta$ are distributed according to a Lorentzian distribution Eq.~\eqref{g},  
the integrals in \eqref{eq:r} and \eqref{eq:v} can then be evaluated   
closing the integral contour in the complex $\eta$-plane, and using
the Cauchy  
residue theorem. Then the firing rate and mean membrane potential   
depend only on the value of $w^{e,i}$ at the pole of $g(\eta)$ in the  
lower half $\eta$-plane: 
$\pi \tau R^{e,i}(\phi,t)+i V^{e,i}(\phi,t) = w^{e,i}(\phi,\bar \eta - i \Delta,t)$,  
and as a result, \eqref{eq:w} must be evaluated only at $\eta=\bar 
\eta-i\Delta$ to obtain the neural field equations (Eq.~\eqref{freG})   
\begin{eqnarray*} 
  \tau \frac{\partial R^{e,i}}{\partial t} &=& \frac{ \Delta }{\pi
                                               \tau}   + 2  R^{e,i}
                                               V^{e,i}, \\  
  \tau \frac{\partial V^{e,i}}{\partial t} & =& \left( V^{e,i}
                                                \right)^2+ \bar \eta -
                                                (\pi \tau R^{e,i})^2+
                                                \tau
                                                S(\phi,t)+P^{e,i}(\phi,t).
\end{eqnarray*} 
where again $S \left( \phi,t \right) = S^e \left( \phi,t \right) + S^i
\left( \phi,t \right)$, and
considering the limit of infinitely fast synapses,  
$\tau_s \to 0$ in Eq.~\eqref{s}, the mean field becomes  
\begin{equation} 
  S^{e,i}(\phi,t)= \frac{1} {2\pi} \int_{-\pi}^{\pi}  J^{e,i}(|\phi-\phi'|)
  R^{e,i}(\phi',t) d \phi',  
  \label{eq:mf0} 
\end{equation} 
Equations \eqref{freG} with  the mean field \eqref{eq:mf0} 
exactly describe the macroscopic dynamics of the population of  
QIF neurons in terms of the local firing rates $R^{e,i}(\phi)$ and mean  
membrane potentials $V^{e,i}(\phi)$. 
These equations can be non-dimensionalized  
by rescaling variables and time as (note the difference between $v^{e,i}_j$, the membrane
potential of a single neuron $j$, and the mean membrane
potential $v^{e,i}$):  
\begin{equation} 
  R^{e,i}= \frac{\sqrt{\Delta}}{\tau}  r^{e,i}, ~V^{e,i} =\sqrt{\Delta} ~ v^{e,i},
  ~t=\frac{\tau}{\sqrt{\Delta}} \tilde t, 
  \label{eq:AdimV} 
\end{equation} 
and parameters as:  
\begin{equation} 
  J^{e,i}_K=\sqrt{\Delta} ~ j^{e,i}_K,~\bar \eta = \Delta~ \tilde \eta,  
  ~P^{e,i}(\phi,t)=\Delta~ \tilde P^{e,i}(\phi,\tilde t). 
  \label{eq:AdimP} 
\end{equation} 
The resulting dimensionless NFM is then 
\begin{subequations} 
  \label{eq:freG} 
  \begin{eqnarray} 
    \dot r^{e,i} &=& \frac{1}{\pi}   + 2  v^{e,i} r^{e,i} ,  
    \label{eq:fre1G}\\ 
    \dot v^{e,i} & =& \left( v^{e,i} \right)^2+ \tilde \eta - \pi^2 \left( r^{e,i} \right)^2+
                      s(\phi,\tilde t) \\
    &&+\tilde P^{e,i}(\phi,\tilde t),\nonumber
    \label{eq:fre2G} 
  \end{eqnarray} 
\end{subequations} 
where the over-dot represents derivation with respect the
non-dimensional time $\tilde t$, and 
the mean field is 
\begin{widetext}
\begin{equation} 
  s(\phi,\tilde{t})=\frac{1} {\pi} \int_{-\pi}^{\pi} \left[\frac{j_0^{e}}{2}+
    \sum_{K=1}^\infty   j_K^{e} \cos (K (\phi'-\phi))\right]
  r^{e}(\phi',\tilde{t}) d \phi'- \frac{1}  {\pi} \int_{-\pi}^{\pi}
  \left[\frac{j_0^{i}}{2}+ \sum_{K=1}^\infty   j_K^{i} \cos (K
    (\phi'-\phi))\right] r^{i}(\phi',\tilde{t}) d \phi'. 
  \label{eq:mfG} 
\end{equation} 
\end{widetext}

\subsection*{Effective NFM model} 
 
Considering $\tilde P^{e,i}(\phi,\tilde t)=\tilde P(\phi,\tilde t)$ in
Eqs.\eqref{eq:freG},  the system    
\begin{subequations} 
  \label{eq:fre} 
  \begin{eqnarray} 
    \dot r &=& \frac{1}{\pi}   + 2  v r ,  
    \label{eq:fre1}\\ 
    \dot v & =& v^2+ \tilde \eta - \pi^2 r^2+ s(\phi,\tilde t) +\tilde
    P(\phi,\tilde t),
    \label{eq:fre2} 
  \end{eqnarray} 
\end{subequations} 
with the mean field  
\begin{equation} 
  s(\phi,t)=\frac{1} {\pi} \int_{-\pi}^{\pi} \left[\frac{j_0}{2}+
    \sum_{K=1}^\infty j_K \cos (K (\phi'-\phi))\right] r(\phi',t) d
  \phi'. 
  \label{eq:mf} 
\end{equation} 
and $$j_K=j_K^e-j_K^i,$$ has identical symmetric solutions as the
original Eqs.\eqref{eq:freG}, i.e.  
$$r^e(t)=r^i(t)=r(t), \quad v^e(t)=v^i(t)=v(t).$$


 \makeatletter 
\setcounter{equation}{0} 
\renewcommand{\theequation}{B\arabic{equation}}
 \makeatother

\section*{Appendix B: Linear stability analysis of the Spatially Homogeneous State} 

\subsection*{Linear stability of effective QIF-NFM Eq.~\eqref{fre}} 

The homogeneous steady state is given by the solution of
Eq.~\eqref{fpr} when $R_{*} \left( \phi \right) = R_{*}$. This is
equivalent to $S_{*} \left( \phi \right) = S_{*} = J_0 R_{*}$ that in dimensionless 
form is 
\begin{equation}
\label{eq:rshs-quartic}
\pi^2r_{*}^4 - j_0 r_{*}^3 -  \tilde \eta r_{*}^2 - \frac{1}{4\pi^2} = 0,
\end{equation}
This equation is greatly simplified assuming $j_0 = 0$, and gives 
\begin{equation}
\label{eq:rshs-j00}
r_{*} = \frac{1}{\pi\sqrt{2}} \sqrt{\tilde \eta  + \sqrt{\tilde{\eta}^2
  + 1}}.
\end{equation}
The stability of homogeneous steady state solutions can be analyzed
studying the evolution of the small (even) perturbations 
($\epsilon \ll 1 $) of the SHS  
\begin{subequations} 
  \label{eq:pert} 
  \begin{eqnarray} 
    r(\phi,t) &=& r_* + \epsilon \sum_{K=0}^\infty a_{K} (t) \cos (K
    \phi), \label{eq:pertr}\\ 
    v(\phi,t) &=& v_* + \epsilon \sum_{K=0}^\infty b_{K} (t) \cos (K
    \phi). \label{eq:pertv} 
  \end{eqnarray} 
\end{subequations} 
Substituting \eqref{eq:pert} into the mean field \eqref{eq:mf}, we  
obtain a perturbed mean field around $s_*(\phi)$ 
\begin{equation}  
  s(\phi,t)=s_*(\phi)+  \epsilon \sum_{K=0}^{\infty} j_K a_K (t)
  \cos(K\phi). 
  \label{eq:Spert} 
\end{equation} 
Linearizing Eqs.~\eqref{eq:fre} around the fixed point $(r_*,v_*)$,
gives
\begin{widetext}
\begin{eqnarray} 
  \sum_{K=0}^\infty \mu_K a_K \cos(K\phi)&=&2 \sum_{K=0}^\infty
  [r_*(\phi) b_K+v_*(\phi) a_K] \cos(K\phi), \nonumber \\ 
  \sum_{K=0}^\infty \mu_K b_K \cos(K\phi)&= & 
  \sum_{K=0}^\infty \left[ 2 v_*(\phi) b_K +(j_K- 2\pi^2 r_*(\phi)  )
    a_K \right] \cos(K\phi) ,  
  \label{eq:linearization} 
\end{eqnarray} 
\end{widetext}
where we have used the Ansatz  
$a_K(t)=a_K e^{\mu_K t}$ and $b_K(t)=b_K e^{\mu_K t}$, where $\mu_K$
represents the dimensionless  eigenvalue of the $K$th mode. 
For SHS states, $(r_*(\phi),v_*(\phi))=(r_{*} ,v_{*} )$, the modes 
in Eqs.~\eqref{eq:linearization} decouple and, for a given mode
$K$, we find the linear system  
\begin{equation}  
  \mu_K 
  \begin{pmatrix}  
    a_K\\ 
    b_K 
  \end{pmatrix} 
  = L_{*}  
  \begin{pmatrix}  
    a_K\\ 
    b_K 
  \end{pmatrix} ,
  \label{eq:linear2} 
\end{equation} 
with: 
\begin{equation}  
  L_{*}=\begin{pmatrix} 
    2 v_{*}  & 2 r_{*} \\ 
    j_K-2 \pi^2 r_{*}  & 2 v_{*}    
  \end{pmatrix} .
  \label{eq:jacobian} 
\end{equation} 
Equation \eqref{eq:linear2} has a general solution: 
\begin{equation}  
  \begin{pmatrix}  
    a_K(\tilde t)\\ 
    b_K(\tilde t) 
  \end{pmatrix} 
  = A_+ 
  \mathbf{u}_+ e^{\mu_{K+} \tilde t} 
  + A_- \mathbf{u}_-  e^{\mu_{K-} \tilde t} , 
  \label{eq:solution_linear} 
\end{equation} 
where $A_{\pm}$ are arbitrary constants. The eigenvalues $\mu_{K\pm}$
are  given by  
\begin{equation} 
  \mu_{K\pm}=
  -\frac{1}{\pi r_*} \pm 2 \pi r_* \sqrt{\frac{j_K}{2 \pi^2  r_* }-1}, 
  \label{eq:mu0} 
\end{equation} 
with eigenvectors  
\begin{equation} 
  \mathbf{u}_\pm= 
  \begin{pmatrix}  
    \pm 1 \\ \sqrt{\frac{j_K }{2r_{*} }- \pi^2} 
  \end{pmatrix}. 
  \label{eq:eigenvect} 
\end{equation} 
In terms of the dimensional variables and parameters (\ref{eq:AdimV}, \ref{eq:AdimP}),  
the eigenvalues \eqref{eq:mu0} are $\lambda_k t = \mu_k
\tilde{t}$, and thus $\lambda_k =\sqrt{\Delta}\mu_k/\tau$. 
In dimensional form, Eq.~\eqref{eq:mu0} is
\begin{equation}
\label{eq:mu1}
  \mu_{K\pm}=-\frac{\sqrt{\Delta}}{\pi \tau R_{*}} \pm \frac{2 \pi \tau R_{*}}{\sqrt{\Delta}}\sqrt{\frac{J_K}{2 \pi^2 \tau R_* }-1}, 
\end{equation}
and the eigenvalues give Eq.~\eqref{lambda} in the main text.

\subsection*{Linear stability of the full QIF-NFM} 
 
For the full QIF-NFM Eq.~\eqref{freG}, the perturbation around the SHS state has the form
\begin{eqnarray} 
  r^{e,i}(\phi,t) &=& r_* + \epsilon \sum_{K=0}^\infty a_{K}^{e,i} (t)
  \cos (K \phi), \nonumber  \\ 
  v^{e,i}(\phi,t)&=& v_* + \epsilon \sum_{K=0}^\infty b_{K}^{e,i} (t)
  \cos (K \phi).  
  \nonumber 
\end{eqnarray} 
In this case, the linear stability of the SHS state with respect to
perturbations of the $K$-spatial  
mode is determined by the characteristic equation
{\small
\begin{equation}  
  \lambda_K 
  \begin{pmatrix}  
    a_K^e\\ 
    b_K^e\\ 
    a_K^i\\ 
    b_K^i 
  \end{pmatrix} 
  =  
  \begin{pmatrix} 
    2 v_{*}  & 2 r_{*}  & 0 & 0 \\ 
    j_K^e-2 \pi^2 r_{*}  & 2 v_{*}  & -j_K^i & 0\\   
    0 & 0 & 2 v_{*}  & 2 r_{*}  \\ 
    j_K^e  & 0 &  -j_K^i-2 \pi^2 r_{*} &2 v_{*}  \\ 
  \end{pmatrix} 
  \begin{pmatrix}  
    a_K^e\\ 
    b_K^e\\ 
    a_K^i\\ 
    b_K^i 
  \end{pmatrix} 
  \label{eq:linear_ei} 
\end{equation}
}
For each $K$ mode, the linearized system has a general solution 
\begin{eqnarray}  
  \begin{pmatrix}  
    a_K^e(\tilde t)\\ 
    b_K^e(\tilde t)\\ 
    a_K^i(\tilde t)\\ 
    b_K^i(\tilde t) 
  \end{pmatrix} 
  =& A_+ & \mathbf{u}_{K+} e^{\mu_{K+} \tilde t} + A_- \mathbf{u}_{K-}
  e^{\mu_{K-} \tilde t} + \nonumber \\
  &B_+& \mathbf{u}_{K\perp} e^{\mu_\perp \tilde t} + B_- \mathbf{\bar
    u}_{K\perp}   
  e^{\bar \mu_\perp \tilde t}, 
  \label{eq:solution_linear_ei} 
\end{eqnarray} 
where $A_{\pm}$ and $B_{\pm}$ are arbitrary constants.  
The eigenvectors   
\begin{equation} 
  \mathbf{u}_{K\pm}= 
  \begin{pmatrix}  
    \pm 1 \\  
    \sqrt{\frac{j_K^e-j_K^i }{2r_{*} }- \pi^2}\\ 
    \pm 1 \\  
    \sqrt{\frac{j_K^e-j_K^i }{2r_{*} }- \pi^2} 
  \end{pmatrix}. 
  \label{eq:eigenvect} 
\end{equation} 
have eigenvalues   
\begin{equation} 
  \mu_{K\pm}=-\frac{1}{\pi r_*} \pm 2 \pi r_*
  \sqrt{\frac{j_K^e-j_K^i}{2 \pi^2  r_* }-1}. 
  \label{eq:mu00} 
\end{equation} 
These eigenvalues coincide with those of the reduced system
\eqref{eq:mu0}, and are associated with  
the standing waves shown in Figure 2. Additionally, the eigenvector
\begin{equation} 
  \mathbf{u}_{K\perp}= 
  \begin{pmatrix}  
    i j_K^i \\  
    \pi j_K^i\\ 
    i j_K^e \\ 
    \pi j_K^e 
  \end{pmatrix}, 
  \label{eq:eigenvect} 
\end{equation} 
and its complex conjugate $\mathbf{\bar u}_{K\perp}$, with associated
eigenvalue  
\begin{equation} 
  \mu_\perp=-\frac{1}{\pi r_*}  +i 2 \pi r_* . 
  \label{eq:muT} 
\end{equation} 
and its complex conjugate $\bar \mu_\perp$, correspond to modes of
oscillation of the uncoupled  system. 
Indeed, note that the eigenvalues \eqref{eq:muT} are
independent of the connectivity,  and correspond to oscillatory  
modes which are already present in a single population of uncoupled
neurons  
---note that eigenvalues \eqref{eq:mu00} reduce to \eqref{eq:muT} for
all the modes  
with $j_K=j_K^e-j_K^e=0$.

\section*{Appendix C: Small-amplitude equation 
near the Spatially Homogeneous State} 

 \makeatletter 
\setcounter{equation}{0} 
\renewcommand{\theequation}{C\arabic{equation}}
 \makeatother

\subsection*{Critical eigenvectors} 
 
Right at the bifurcation, the only undamped mode is the critical one
given by  
$\mathbf{u}_+$ in \eqref{eq:eigenvect}, that reduces to the critical
eigenmode:  
\begin{equation} 
  \mathbf{u}_{c}=\binom{r_*}{-v_*}. 
  \label{eq:cr_eigenmode} 
\end{equation}  
At criticality, the critical eigenmode of $L_{*}$ satisfies  
$$ L_{*c} \mathbf{u}_{c}=0$$  
where $L_{*c}$ corresponds to the operator \eqref{eq:jacobian}
evaluated  
at $j_K=j_{Kc}$. The left critical eigenvector of the operator
$L_{*c}$ is 
then defined as  
$$ \mathbf{u}_{c}^\dagger L_{*c} =0$$  
what gives   
\begin{equation} 
  \mathbf{u}_{c}^\dagger=\pi\binom{-v_*}{r_*}^T, 
  \label{eq:lne} 
\end{equation}  
where the constant has been taken  
to normalize the eigenvectors, so that they satisfy  
$\mathbf{u}_{c}^\dagger \mathbf{u}_{c}=1$.

\subsection*{Amplitude equation} 
 
Except for initial transients, the amplitude of the bifurcating
solution  
at criticality is expected to contain only the component
$\mathbf{u}_{+c}$.           
In the following we derive a small-amplitude equation for the bump
solutions    
using multiple-scale analysis, see e.g. \cite{Kur84}.  
First, let the solution of Eqs.~\eqref{eq:fre} be written as the  
perturbation expansion  
\begin{equation} 
  \binom{r(\phi,\tilde t)}{v(\phi,\tilde t)}=\binom{r_{*} }{v_{*} }+ 
  \epsilon \binom{r_\epsilon (\phi,\tilde t,\tilde{T})}{v_{\epsilon}
    (\phi,\tilde t,\tilde{T})}+ 
  \epsilon^2 \binom{r_{\epsilon \epsilon} (\phi,\tilde
    t,\tilde{T})}{v_{\epsilon \epsilon} (\phi,\tilde
    t,\tilde{T})}+\dots 
  \label{eq:exp} 
\end{equation} 
where $(r_{*} ,v_{*} )$ is the state SHS given by the solutions of
\eqref{eq:rshs-quartic},  
and $\epsilon \ll 0$ is a small parameter, which measures the distance
from the  
Turing bifurcation. In addition we define a long time scale
$\tilde{T}=\epsilon^2 \tilde t$, that 
is considered to be  independent of $\tilde t$.  Accordingly, the
differential  
operator in Eqs.~\eqref{eq:fre} may be replaced by:  
$$\partial_{\tilde t} \to \partial_{\tilde
  t}+\epsilon^2 \partial_{\tilde{T}}.$$ 
Since the asymptotic expansion is going to be performed in the
vicinity  
of a stationary bifurcation, we set $\partial_{\tilde t}=0$ so that 
the only temporal variations occur with the slow time scale 
$\tilde{T}$.
 
Additionally, in our analysis we use the parameter $j_1$ as the
bifurcation parameter, and we   
write it as  
\begin{equation} 
  \label{eq:j1c} 
  j_1=j_1^T+  \epsilon^2 \delta j_1,  
\end{equation} 
where $j_1^T$ is the critical value of $j_1$ at which the Turing  
bifurcation occurs, given by Eq.~\eqref{Jc}, with $K=1$.  
Accordingly, the (non-dimensionalized) connectivity footprint
\eqref{J} is 
\begin{equation} 
  j(\phi)=j_c(\phi)+2 \epsilon^2 \delta j_1 \cos\phi , 
  \label{eq:expJ} 
\end{equation} 
with  
\begin{equation} 
  j_c(\phi)=j_0+2 j_1^T \cos\phi+2 \sum_{K=2}^\infty j_K\cos(K \phi),  
  \label{eq:defJc} 
\end{equation}  
where $j_K<j_{Kc}$ for $K\neq1$. 
To simplify the notation, we hereafter omit to explicitly write  
the dependence of $r_{\epsilon,\epsilon \epsilon,\dots}$ and
$v_{\epsilon,\epsilon \epsilon,\dots}$ on the variables $\tilde t,T$
and $\phi$. 
Substituting \eqref{eq:exp} and \eqref{eq:expJ} into the mean field
\eqref{eq:mf}:  

\begin{widetext}

\begin{eqnarray} 
  s(\phi)&=&\frac{1}{2\pi} \int_{-\pi}^\pi(r_{*} +\epsilon r_\epsilon
  +\epsilon^2 r_{\epsilon \epsilon} +\dots)  j_c(\phi-\phi') d\phi'+ 
  \epsilon^2 \frac{1}{\pi} \int_{-\pi}^\pi (r_{*} +\epsilon r_\epsilon
  +\epsilon^2 r_{\epsilon \epsilon} +\dots) ~ \delta j_1
  \cos(\phi-\phi') d\phi'  \nonumber  \\
  & \equiv & \langle r_{*} +\epsilon r_\epsilon +  
  \epsilon^2 r_{\epsilon \epsilon} +\dots \rangle_c + 
  2 \epsilon^2 \langle r_{*} +\epsilon r_\epsilon +\epsilon^2
  r_{\epsilon \epsilon} +\dots  \rangle  
  \label{eq:def} \\ 
  &=& r_{*}  j_0 +\epsilon \langle r_\epsilon \rangle_c  
  +\epsilon^2  \langle r_{\epsilon \epsilon}  \rangle_c+ 
  \epsilon^3 ( \langle r_{\epsilon \epsilon \epsilon}  \rangle_c+
  2\langle r_\epsilon  \rangle)+\dots 
  \label{eq:scalarP} 
\end{eqnarray} 
Plugging expansions \eqref{eq:exp} and \eqref{eq:expJ} into the NFM
Eqs.~\eqref{eq:fre},  we obtain 
\begin{eqnarray} 
  \epsilon^{2} \partial_{\tilde{T}}( 
  \epsilon r_\epsilon +\epsilon^2 r_{\epsilon \epsilon}
  +\dots)&=&\nonumber 
  \epsilon (2 v_{*}  r_\epsilon  + 2 r_{*}  v_{\epsilon} ) +
  \epsilon^2 (2 v_{*}  r_{\epsilon \epsilon}  + 2 r_\epsilon
  v_{\epsilon}  +  
  2 r_{*}  v_{\epsilon \epsilon} ) +  
  \epsilon^3  (2  v_{\epsilon}  r_{\epsilon \epsilon} + 2 r_\epsilon
  v_{\epsilon \epsilon} ) + \dots \nonumber 
  \\  
  \epsilon^{2} \partial_{\tilde{T}}( 
  \epsilon v_{\epsilon} +\epsilon^2 v_{\epsilon \epsilon}  +\dots)&=& 
  \epsilon  (2 v_{*}  v_{\epsilon}  - 2\pi^2 r_{*}  r_\epsilon  +
  \langle r_\epsilon \rangle_c)  +\epsilon^2 ( v_{\epsilon} ^2-\pi^2
  r_\epsilon ^2  + 2 v_{*}   v_{\epsilon \epsilon} - 2 \pi^2 r_{*}  r_{\epsilon \epsilon}   + \langle r_{\epsilon \epsilon}  \rangle_c )  +\nonumber \\
  && \epsilon^3 (2 v_{\epsilon}  v_{\epsilon \epsilon}-2 \pi^2
  r_\epsilon  r_{\epsilon \epsilon} + \langle r_{\epsilon \epsilon
    \epsilon}  \rangle_c+2 \langle r_\epsilon    \rangle ) + \dots \nonumber 
\end{eqnarray} 
These equations can be written in a more compact form as 
\begin{equation} 
  -(L_c+\epsilon^{2} L_{\epsilon \epsilon})\left[ \epsilon 
    \binom{r_\epsilon }{v_{\epsilon} }+ 
    \epsilon^2 \binom{r_{\epsilon \epsilon} }{v_{\epsilon \epsilon} }+ 
    ...\right] = \epsilon^{2}N_{\epsilon \epsilon} + 
  \epsilon^{3}N_{\epsilon \epsilon \epsilon} +\dots,\label{eq:MSeq} 
\end{equation} 
\end{widetext}

defining the linear and nonlinear operators 
\begin{eqnarray} 
  L_c &=&  
  \begin{pmatrix} 
    2 v_{*}  & 2 r_{*} \\ 
    \langle \cdot \rangle_c -2 \pi^2 r_{*}  & 2 v_{*}    
  \end{pmatrix}, 
  \nonumber \\ 
  L_{\epsilon \epsilon} &=&  
  \begin{pmatrix} 
    -\partial_{\tilde{T}} & 0\\ 
    2\langle \cdot \rangle & -\partial_{\tilde{T}}    
  \end{pmatrix}, 
  \nonumber \\  
  N_{\epsilon \epsilon}  &=& 
  \binom{2 r_\epsilon  v_{\epsilon} }{v_{\epsilon} ^2 -\pi^2
    r_\epsilon ^2}, 
  \nonumber \\  
  N_{\epsilon \epsilon \epsilon}  &=&  \binom{2r_\epsilon  v_{\epsilon
      \epsilon} +2r_{\epsilon \epsilon}  v_{\epsilon} }{2v_{\epsilon}
    v_{\epsilon \epsilon} -2\pi^2 r_\epsilon  r_{\epsilon \epsilon} },\nonumber
  \nonumber 
\end{eqnarray} 
Next we collect terms by order in $\epsilon$. At first order we
recover the  
linear problem \eqref{eq:linear2} at the Turing bifurcation: 
\begin{equation} 
  \begin{pmatrix} 
    2 v_{*}  & 2 r_{*} \\ 
    j_1^T-2 \pi^2 r_{*}  & 2 v_{*}    
  \end{pmatrix} 
  \binom{r_\epsilon }{v_{\epsilon} }=\binom{0}{0}. 
  \nonumber 
\end{equation} 
Recalling that $j_1^T$ is given by Eq.~\eqref{eq:j1c}, we find the 
neutral solution:  
\begin{equation} 
  \binom{r_\epsilon }{v_{\epsilon} }=A~\mathbf{u}_c \cos \phi, 
  \label{eq:O1s} 
\end{equation} 
where $A$ is the small amplitude with slow time dependence that   
we aim to determine, and $\mathbf{u}_c$ is the critical eigenmode   
given by Eq.~\eqref{eq:cr_eigenmode}.  
Substituting the solution \eqref{eq:O1s} into  
the nonlinear forcing terms $N_{\epsilon \epsilon}$ we find  
\begin{equation} 
  N_{\epsilon \epsilon}  = \frac{A^2}{2}\binom{\pi^{-1}}{v_{*}
    ^2-\pi^2 r_{*} ^2} [1+\cos(2\phi)], 
  \nonumber  
\end{equation} 
what implies that, at second order, the solution must necessarily  
contain homogeneous and   
second spatial components   
\begin{equation} 
  \binom{r_{\epsilon \epsilon} }{v_{\epsilon \epsilon} } =
  \binom{r_{\epsilon \epsilon 0}}{v_{\epsilon \epsilon
      0}}+\binom{r_{\epsilon \epsilon 2}}{v_{\epsilon \epsilon  2}}\cos(2\phi).
  \nonumber 
\end{equation} 
Equating the homogeneous, second order terms of equation
\eqref{eq:MSeq} we find   
\begin{equation} 
  - 
  \begin{pmatrix} 
    2 v_{*}  & 2 r_{*} \\ 
    j_0-2 \pi^2 r_{*} & 2 v_{*}    
  \end{pmatrix}  
  \binom{r_{\epsilon \epsilon 0}}{v_{\epsilon \epsilon 0}} =  
  \frac{A^2}{2}\binom{\pi^{-1}}{v_{*} ^2-\pi^2 r_{*} ^2} , \nonumber 
\end{equation} 
and left-multiplying this equation by $L_c^{-1}$, and using
Eq.~\eqref{JT} we find  
\begin{equation} 
  \binom{r_{\epsilon \epsilon 0}}{v_{\epsilon \epsilon 0}} =  
  \frac{A^2}{4 r_*(j_1^T-j_0)}     
  \begin{pmatrix} 
    2 v_{*}  & -2 r_{*} \\ 
    2 \pi^2 r_{*}-j_0 & 2 v_{*}    
  \end{pmatrix} 
  \binom{\pi^{-1}}{v_{*} ^2-\pi^2 r_{*} ^2} ,\nonumber 
\end{equation} 
what gives the coefficients  
\begin{eqnarray} 
  r_{\epsilon \epsilon 0} &=&\frac{3v_{*} ^2-\pi^2r_{*}
    ^2}{2(j_1^T-j_{0})}  A^2, \label{eq:r20}\\ 
  v_{\epsilon \epsilon 0} &=&\frac{2\pi v_{*} ^4-v_{*}
    j_{0}-3\pi/2}{2(j_1^T-j_{0})}  A^2. 
  \label{eq:v20} 
\end{eqnarray} 
Proceeding similarly, we find the coefficients corresponding to the
second spatial  
Fourier modes: 
\begin{eqnarray} 
  r_{\epsilon \epsilon 2} &=&\frac{3v_{*} ^2-\pi^2r_{*}
    ^2}{2(j_1^T-j_{2})}  A^2, 
  \label{eq:r22}\\ 
  v_{\epsilon \epsilon 2} &=&\frac{2\pi v_{*} ^4-v_{*}
    j_{2}-3\pi/2}{2(j_1^T-j_{2})}  A^2. 
  \label{eq:v22} 
\end{eqnarray} 
Collecting the third order terms of equation \eqref{eq:MSeq} we obtain
the identity 
\begin{equation} 
  -L_c \binom{r_{\epsilon \epsilon \epsilon} }{v_{\epsilon \epsilon
      \epsilon}} -L_{\epsilon \epsilon} \binom{r_\epsilon
  }{v_{\epsilon} }= N_{\epsilon \epsilon \epsilon} , 
  \label{eq:O3eq} 
\end{equation} 
To obtain the desired amplitude equation, we shall left-multiply
Eq.~\eqref{eq:O3eq} 
by the left null-eigenvector \eqref{eq:lne} 
and project it into the first spatial Fourier mode.  
The first term on the r.h.s. of Eq.~\eqref{eq:O3eq} vanishes since  
$ \mathbf{u}_c^\dagger L_c=0$. The second term is    
\begin{equation} 
  L_{\epsilon \epsilon}  \binom{r_\epsilon }{v_{\epsilon} }=  \binom{-
    r_{*} ~\partial_{\tilde{T}} A} 
  {v_{*} ~\partial_{\tilde{T}} A+\delta j_1~ r_{*}  ~A} \cos \phi.  
  \nonumber 
\end{equation} 
Finally, the nonlinear forcing term at the l.h.s. of Eq.~\eqref{eq:O3eq} is:  
\begin{widetext} 
  \begin{eqnarray} 
    N_{\epsilon \epsilon \epsilon} = -A \cos \phi \binom{v_{*}  (2
      r_{\epsilon \epsilon 0}  + r_{\epsilon \epsilon 2})-r_{*}  (2
      v_{\epsilon \epsilon 0} +  v_{\epsilon \epsilon  2})  }
    {\pi^2 r_{*}  (2 r_{\epsilon \epsilon 0}  +  r_{\epsilon \epsilon
        2}) +v_{*} (2  v_{\epsilon \epsilon 0}+   v_{\epsilon \epsilon
        2})}- 
    A \cos(3 \phi) \binom{  v_{*}  r_{\epsilon \epsilon 2}-r_{*}
      v_{\epsilon \epsilon 2} } 
    {  \pi^2  r_{*}   r_{\epsilon \epsilon 2}+v_{*}  v_{\epsilon
        \epsilon 2} }. 
    \nonumber 
  \end{eqnarray} 
  Thus, the solvability condition gives  
  \begin{eqnarray} 
    \mathbf{u}_c^\dagger  \binom{r_{*} ~ \partial_{\tilde{T}}
      A}{-v_{*} ~ \partial_{\tilde{T}} A 
      -\delta j_1~ r_{*}  ~A}=   
    -A \mathbf{u}_c^\dagger \binom{v_{*}  (2 r_{\epsilon \epsilon 0}
      + r_{\epsilon \epsilon 2})-r_{*}  (2  v_{\epsilon \epsilon 0} +
      v_{\epsilon \epsilon 2})  } 
    {\pi^2 r_{*}  (2 r_{\epsilon \epsilon 0}  +  r_{\epsilon \epsilon
        2}) +v_{*} (2  v_{\epsilon \epsilon 0}+   v_{\epsilon \epsilon
        2})} . 
    \label{eq:sc} 
  \end{eqnarray} 
  Substituting the coefficients
  (\ref{eq:r20},~\ref{eq:v20},~\ref{eq:r22},~\ref{eq:v22})  
  into Eq.~\eqref{eq:sc} gives the desired amplitude equation 
  \begin{equation} 
    \label{eq:AE} 
    \partial_{\tilde{T}} A= \pi r_{*} ^2 ~\delta j_1 A+ \tilde{a} A^3,  
  \end{equation} 
  where the parameter $a$ is  
  \begin{eqnarray} 
    \tilde{a} =\pi \left( 5 v_*^4 +\pi^4r_{*} ^4 - \frac{5}{2} \right)  
    \left( \frac{1}{j_1^T-j_0}+\frac{1/2}{j_1^T-j_2} \right) 
    - v_{*} \left( \frac{j_0}{j_1^T-j_0}+\frac{j_2/2}{j_1^T-j_2}
    \right). 
    \label{eq:b} 
  \end{eqnarray} 
  Equating Eq.~\eqref{eq:b} to zero, gives the critical boundary
  $j_2^c$  
  separating sub-critical and super-critical Turing bifurcations: 
  \begin{equation} 
    j_2^c=\frac{3 j_1^T-j_0}{2}+\frac{6(j_1^T-j_0)^2 \pi^2 r_{*}
      ^3}{5 + 4 \pi^2 r_{*} ^3  
      (3 j_0 - j_1^T - 10 \pi^2 r_{*}  + 4 \pi^6 r_{*} ^5)} 
    \label{eq:J2c} 
  \end{equation} 

In dimensional form, Eqs.~(\ref{eq:AE},~\ref{eq:b},~\ref{eq:J2c}) are respectively:
\begin{equation} 
  \tau \partial_T A  =  \pi \frac{\tau^2 R_{*}^2}{\Delta} ~\delta 
  J_1 A+ a A^3, 
\label{AmplEq} 
\end{equation} 
\begin{equation} 
  \label{eq:br} 
  a  =   \left[ \pi \left( \frac{5 \Delta^3}{16 \pi^4 \tau^4 R_{*}^4}+
      \frac{\pi^4 
        \tau^4 R_{*}^4}{\Delta} -\frac{5\Delta}{2} \right)\left( 
      \frac{1}{J_1^T- J_0} + 
      \frac{1/2}{J_1^T-J_2} \right) + 
    \frac{\Delta}{2\pi\tau R_{*}} \left( 
      \frac{J_0}{J_1^T-J_0}+\frac{J_2/2}{J_1^T-J_2} \right) \right], 
\end{equation} 
and 
\begin{equation} 
  \label{eq:J2cr} 
  J_2^c  =  \frac{3 J_1^T-J_0}{2}+\frac{6(J_1^T-J_0)^2 \pi^2 
    \tau^3 R_{*}^3}{5 \Delta^2 + 4 \pi^2 \tau^3 R_{*}^3  
    \left(3 J_0 - J_1^T - 10 \pi^2 \tau R_{*}  + \frac{4 \pi^6 \tau^5
      R_{*}^5}{\Delta^2} \right) }. 
\end{equation} 
\end{widetext}

\makeatletter 
\setcounter{equation}{0} 
\renewcommand{\theequation}{D\arabic{equation}}
\makeatother

\section*{Appendix D: Numerical simulations} 
 
\subsection*{Numerical simulation of the QIF model}

\begin{figure}[htbp]            
\label{alg:qif}
  \begin{algorithm}[H]
    \begin{algorithmic}[1]
      \Require Variables: $v_j$,  $I_j$, $t^r_j$ (exit time from refractory
      period), $t$ (time). Constants: $\tau, dt,
      v_p$.
      \Ensure  $\dot{v}_j = v_j^2 + I_j$ and $t_j^l$ and $t^r_j$.
      \State {\textbf{bool} spike$_j \leftarrow $ \textbf{False}}
      \If{$t \geq t^r_j$} \PComment{0.4}{Check whether the neuron is in the
        refractory period.}
      \State{$v_j \leftarrow v_j + \frac{dt}{\tau} \left( v_j^2 + I_j
        \right)$}  \PComment{0.4}{Euler integration.}
      \If{$v_j \geq v_{p}$} \PComment{0.4}{Check if the voltage has crossed the
        threshold.}
      \State spike$_j \leftarrow$ \textbf{True}  \PComment{0.4}{The neuron has spiked at
        time $t_j^{l}$.}
      \State \label{state:refr_period} $t^r_j \leftarrow t +
      2\cdot\frac{\tau}{v_j}$ \PComment{0.4}{Set the end of the refractory period.}
      \State \label{state:spike_time} $t_j^l \leftarrow t +
      \frac{\tau}{v_j}$  \PComment{0.4}{Spike time is set after $\frac{\tau}{v_j}$.}
      \State $v_j \leftarrow -v_j$  \PComment{0.4}{Reset the voltage.}
      \EndIf
      \EndIf
\end{algorithmic}
\end{algorithm}
\caption{Algorithm used for the Euler integration of the QIF neuron~Eq.\eqref{qif1}.}
\end{figure}

In numerical simulations we used the Euler scheme  with time step $dt = 10^{-3}$.
Additionally, we considered the peak and reset values  $v_p = -v_r = 100$.
The Algorithm used to simulate the QIF neuron \eqref{qif1} is shown 
in Fig.~\ref{alg:qif}.

\subsection*{Numerical simulation of the ring network}

To numerically implement the ring network of QIF neurons we
divided the ring into $n=100$ intervals located at $\phi_m =2\pi
m/n-\pi$, $m = 1, \dots, n$. 
At each interval $\phi_m$, we considered $N^e_m=N/(2n) = 2.5\cdot
10^3$ excitatory and $N^i_m=2.5\cdot 10^3$ inhibitory neurons. 

Then we distributed the neurons in each interval $\phi_m$ using a
Lorentzian distribution Eq.~\eqref{g}. For each $\phi_m$ the we used the inverse cumulative
 distribution function (quantile function):
\begin{equation}
\label{eq:1}
\eta_i = \bar \eta + \Delta \tan \left[ \frac{\pi}{2} \frac{2i - N_m -
  1}{N_m + 1} \right],\ i = 1, \dots, N_m.
\end{equation}

Perturbations were implemented such that,
at a certain time $t_0 = 0.05$ s, a spatially modulated pulse was
applied with the form:
\begin{equation}
\label{eq:perturbations}
P^{e,i} \left( \phi, t \right) = A \left(e^{\left( t-t_0 \right)/\tau_r} -1\right)\cdot\cos
\left( K\cdot \phi \right), 
\end{equation}
where the amplitude was $A = 0.3$, $K$ represents the wavenumber of the perturbation and
$\tau_r = 4\cdot 10^{-3}$ s  is the rising time constant of the perturbation. The
perturbations had a duration of $\Delta t = 0.01$ s. 
\newline

The instantaneous firing rates in Fig. \ref{Figure2} are obtained
binning time and counting the spikes of neurons in each interval $\phi_{m}$  within a sliding time window of size
$\delta t = 0.01$s (in dimensionless time, $\delta \tilde{t} = 0.5)$. 

\end{document}